\documentclass[aps,prl,reprint,superscriptaddress,longbibliography,floatfix]{revtex4-2}
\usepackage{physics,graphicx,amssymb,bm,amsmath,xcolor,braket,microtype,hyperref,siunitx}
\hypersetup{colorlinks=true,linkcolor=blue,urlcolor=blue,citecolor=blue}

\def\v{{\mathbf v}}
\def\sig{{\boldsymbol\sigma}}

\def\ev{{\mathbf E}}
\def\rv{{\mathbf r}}
\def\q{{\mathbf q}}
\def\n{{\mathbf n}}
\def\m{{\mathbf m}}

\newcommand{\CMcal}[1]{\mathcal{#1}}

\begin{document}

\title{Nonlinear Topological Orbital Responses of Antiferromagnetic Skyrmions}

\author{Amir N. Zarezad}
\affiliation{Department of Physics, Institute for Advanced Studies in Basic Sciences (IASBS), Zanjan 45137-66731, Iran}
\author{Arne Brataas}
\affiliation{Center for Quantum Spintronics, Department of Physics, Norwegian University of Science and Technology, NO-7491 Trondheim, Norway}
\author{Alireza Qaiumzadeh}
\affiliation{Center for Quantum Spintronics, Department of Physics, Norwegian University of Science and Technology, NO-7491 Trondheim, Norway}

\date{\today}

\begin{abstract}
Antiferromagnetic skyrmions evade the skyrmion Hall effect, but compensation suppresses their conventional topological charge Hall signal. We predict a semiclassical nonlinear topological orbital response of an isolated skyrmion in a $\mathcal{PT}$-symmetric hexagonal antiferromagnet without spin-orbit coupling. In the diffusive, weak-emergent-field regime, the spin-dependent emergent Lorentz force reshapes the carrier distribution, generating a local orbital Hall-current correction and a local orbital accumulation, both quadratic in the applied electric field. The current correction requires spin-asymmetric longitudinal scattering, whereas the accumulation survives spin-symmetric scattering. When the spin-diffusion length greatly exceeds the strip width, the current correction persists while the accumulation approaches zero. Both signals are even under electric-field reversal, odd under reversal of the skyrmion topological charge $Q$, and helicity independent within this model. These symmetries enable rectified detection and distinguish this mechanism from the $Q$-even quantum-regime topological orbital Hall effect.
\end{abstract}

\maketitle
Topological magnetic solitons, particularly magnetic skyrmions, are localized spin textures with particle-like properties that exhibit  stability against external perturbations \cite{Polyakov,bogdanov1989thermodynamically,bogdanov1994thermodynamically,doi:10.1126/science.1166767}. Their topological protection and current-driven dynamics make them promising building blocks for next-generation spintronic devices, including racetrack memories and logic gates~\cite{Nagaosa2013, Fert2017, Nayak2017}. In conducting ferromagnets, the real-space topology acts as an emergent magnetic gauge field for itinerant electrons, giving rise to the topological charge Hall effect (TCHE), which provides a direct electrical signature of skyrmions~\cite{PhysRevLett.93.096806,PhysRevB.97.134401}. The same topology, however, also produces the skyrmion Hall effect, where the Magnus force deflects moving skyrmions toward the sample edges and may lead to their annihilation~\cite{Nagaosa2013, Jiang2016, Litzius2017}.

Early continuum theories predicted localized antiferromagnetic
(AFM) skyrmions in noncentrosymmetric
crystals~\cite{bogdanov1998vortex,PhysRevB.66.214410}.
Experiments have demonstrated isolated skyrmions in synthetic
AFMs~\cite{juge2022skyrmions}, fractional AFM skyrmion lattices
in MnSc$_2$S$_4$~\cite{gao2020fractional}, and merons and bimerons
in insulating $\alpha$-Fe$_2$O$_3$~\cite{jani2021antiferromagnetic}.
Related studies of centrosymmetric metals have identified
field-induced skyrmion lattices in
Gd$_2$PdSi$_3$~\cite{doi:10.1126/science.aau0968} and linked
helical order to the electronic structure of
GdRu$_2$Si$_2$~\cite{eremeev2023insight}.
Isolated, fully compensated skyrmions in intrinsic metallic AFMs
remain an experimental goal, offering the prospect of motion
without the skyrmion Hall
effect~\cite{Ezawa2016,Zhang2016,PhysRevLett.116.147203,
PhysRevB.99.054423}. Because the two magnetic sublattices carry opposite topological charges, the Magnus forces cancel, allowing for straight and potentially ultrafast motion without the skyrmion Hall effect~\cite{PhysRevB.96.060406,  Ezawa2016,PhysRevLett.116.147203}. This compensation, however, also suppresses the conventional TCHE, making skyrmions in AFM systems with parity-time ($\mathcal{PT}$) symmetry difficult to detect using standard charge-transport measurements. Although the opposite emergent magnetic fields acting on opposite spin species generate a finite topological spin Hall effect (TSHE)~\cite{Buhl.pssr.2017, PhysRevLett.121.097204, PhysRevB.110.054431,Zarezad_2024}, offering a route to pure spin-current generation in AFM skyrmions, complementary transport phenomena that exploit other electronic degrees of freedom remain largely unexplored.

\begin{figure}[t]\label{schematic}
    \centering
\includegraphics[width=\columnwidth]{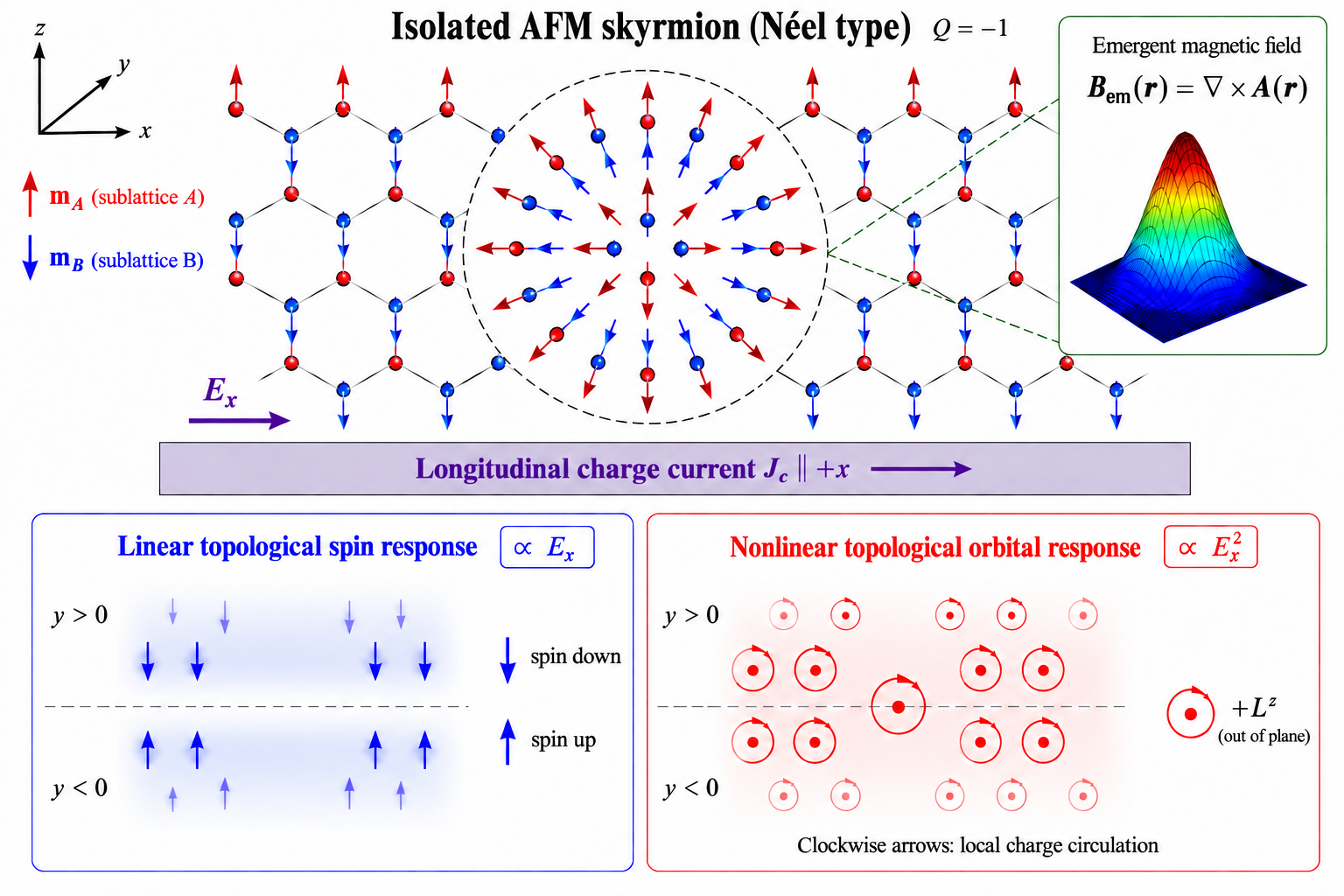}
    \caption{Schematic illustration of the topological spin and nonlinear topological orbital responses of an isolated Néel skyrmion in a hexagonal antiferromagnet. A longitudinal electric field drives a charge current, while the skyrmion’s emergent magnetic field \(\mathcal{B}_{\rm em}\) generates distinct transverse responses. Bottom Left: the linear topological spin Hall effect produces a spin Hall current and an antisymmetric spin accumulation. The spin accumulation therefore vanishes at \(y=0\), has opposite signs on the two sides of the skyrmion, and reaches extrema at finite \(|y|\). Its detailed spatial profile, including its values at the strip edges, depends on the spin-diffusion length, as shown in Fig. \ref{fig:results}(a). Bottom Right: the nonlinear topological orbital response comprises a skyrmion-induced correction to the orbital Hall current and a local nonequilibrium orbital polarization, both proportional to \(E_x^2\). When finite, the orbital polarization is even in \(y\), has the same sign on both sides of the skyrmion, and reaches its maximum at \(y=0\), consistent with Fig. \ref{fig:results}(b). Reversing the skyrmion topological charge \(Q\) reverses both the linear topological spin and nonlinear topological orbital responses, whereas reversing \(E_x\) reverses only the linear spin response. This schematic was generated with OpenAI ChatGPT (GPT-5) under author direction and scientifically verified by the authors.}
    \label{fig:schematic}
\end{figure}

Alongside spin transport, the orbital degree of freedom has recently emerged as a distinct and versatile carrier of information in solids~\cite{PhysRevLett.94.066602, PhysRevLett.95.066601, PhysRevLett.100.096601}. 
Experiments have demonstrated long-range orbital transport and large orbital torques~\cite{PhysRevB.107.134423, Hayashi2023}. Even in centrosymmetric crystals, an applied electric field can drive a transverse orbital Hall current through interband coherence associated with momentum-space orbital textures, without requiring spin-orbit coupling or spin polarization~\cite{Go_2021, PhysRevB.98.214405, PhysRevLett.121.086602}. 
This transport response is distinct from the orbital Edelstein effect, in which a longitudinal current generates a spatially uniform orbital polarization and which generally requires broken inversion symmetry at linear order~\cite{Salemi2019, PhysRevResearch.3.013275}.
More recently, nonlinear orbital Hall currents and the associated out-of-plane orbital torques have been proposed in noncentrosymmetric topological materials~\cite{wang2025outofplanenonlinearorbitalhall} and \(\mathcal{PT}\)-symmetric antiferromagnets with centrosymmetric crystal structures \cite{NEW_NONLINEAR_OHE_REF}, while current-induced orbital magnetization has been linked to nonlinear charge Hall responses~\cite{42zw-rs43}.

A distinct mechanism for orbital transport arises from real-space magnetic topology. Noncollinear magnetic textures can induce orbital responses through spin chirality, even in the absence of spin-orbit coupling~\cite{Dias2016, Lux2018}, leading to the topological orbital Hall effect (TOHE)~\cite{Goebel2025}. 
Previous work on the skyrmion-induced TOHE focused on the strong-coupling quantum regime of nanometric skyrmion crystals, where the emergent magnetic field produces Landau quantization and orbital-polarized edge currents~\cite{Goebel2025}, building on related quantum topological Hall physics~\cite{PhysRevB.92.115417, PhysRevB.95.094413}. Here, we instead address the weak-field diffusive regime, in which the texture modifies the nonequilibrium carrier distribution without quantizing the electronic spectrum.

In this Letter, we demonstrate that an isolated AFM skyrmion generates a nonlinear TOHE in the semiclassical transport regime; see Fig. \ref{fig:schematic}. The nonlinear response originates from the coupling between the emergent magnetic field of the AFM skyrmion and the nonequilibrium carrier distribution. This mechanism produces an orbital Hall current quadratic in the applied electric field, in contrast to the linear TSHE driven by the same emergent field. We derive analytical expressions for both the nonlinear topological orbital Hall conductivity and the induced orbital accumulation. 
When the spin-diffusion length is comparable to or shorter than the transverse strip width, the orbital accumulation exhibits a pronounced spatial profile that encodes the skyrmion texture. In the opposite limit, the gradient-mediated and direct emergent-field contributions cancel, and the orbital accumulation vanishes.
Our theoretical framework combines the low-energy Dirac theory of a hexagonal AFM with semiclassical Boltzmann transport, the Kubo formalism, and the modern theory of orbital magnetization \cite{RevModPhys.82.1959}. 

\textit{Model and emergent gauge field.} We consider a $\mathcal{PT}$-symmetric AFM on a two-dimensional hexagonal lattice with sublattices $A$ and $B$, and a lattice constant $a_0$. The tight-binding Hamiltonian is \cite{PhysRevB.110.054431}
\begin{equation} 
\begin{aligned}
\mathcal{H} =&-t\sum_{\rv \in A}\sum_{i=1}^{3}\sum_{\sigma}
\Big[a_{\sigma}^{\dagger}(\rv)b_{\sigma}(\rv+\bm{\xi}_i)+\mathrm{H.c.}\Big] \\
&-J\sum_{\sigma\sigma'}
\Bigg[
\sum_{\rv\in A}\m_a\cdot\sig_{\sigma\sigma'}
a_{\sigma}^{\dagger}a_{\sigma'}
+
\sum_{\rv\in B}\m_b\cdot\sig_{\sigma\sigma'}
b_{\sigma}^{\dagger}b_{\sigma'}
\Bigg],
\end{aligned}
\label{eq:H}
\end{equation}
where $a$ ($a^\dagger$) and $b$ ($b^\dagger$) annihilate (create) electrons on the $A$ and $B$ sublattices, respectively, $\bm{\xi}_i$ are the nearest-neighbor vectors, $t$ is the hopping amplitude, $J$ is the $sd$ exchange coupling between the itinerant electrons and the localized magnetic moments, $\bm{m}_{a(b)}$
is the unit vector along the $A$ $(B)$ sublattice moment, and \(\boldsymbol\sigma\) denotes the spin Pauli matrices.

Magnetic interactions stabilizing skyrmion textures can be tuned by carrier doping~\cite{PhysRevLett.120.197202} or light~\cite{PhysRevB.100.060410,PhysRevB.103.134428}. Here, we treat the skyrmion as a prescribed slowly varying texture and isolate its effect on electronic transport.

In the collinear ground state, $\m_a=-\m_b=\hat{z}$, we define the staggered Néel vector as $\n=(\m_a - \m_b)/2$. Linearizing the spectrum around the two inequivalent Dirac points, $K_\pm=\pm \left(4\pi/(3\sqrt{3}a_0),0\right)$, yields $\varepsilon_\eta(\q)=
\eta\sqrt{J^2+|\gamma_\q|^2}$,
where $\eta=\pm1$ labels the conduction and valence bands, $\gamma_\q \simeq \hbar v_0(q_x\pm iq_y)$ is the structure factor around the $K_\pm$ points, and $v_0=3ta_0/2\hbar$ is the massless Dirac Fermi velocity. The low-energy Hamiltonian has the same form as that of gapped graphene. However, unlike graphene, where the gap arises from a staggered sublattice potential, here it is generated by the AFM exchange interaction $J$.
The corresponding eigenstates $\Psi^{\mathrm{s}}_{\eta}$ are given in the Supplemental Material (SM)~\cite{SM}. Owing to the combined $\mathcal{PT}$ symmetry, all bands remain spin degenerate. The Hamiltonian (\ref{eq:H}) is equally applicable to synthetic AFMs with antiferromagnetically coupled magnetic layers.

We consider an isolated AFM skyrmion embedded in a uniform collinear AFM background, such that $\mathbf n\to\hat{z}$ as $r\to\infty$. 
The Néel vector is parameterized as \(\mathbf n(\mathbf r)=(\cos\Phi\sin\Theta,\sin\Phi\sin\Theta,\cos\Theta)\), with \(\Theta(r)=2\pi-4\arctan[\exp(4r/r_0)]\) and \(\Phi(\varphi)=\varsigma\varphi+\gamma\), where \(r_0\) sets the skyrmion size, \(\varsigma=\pm1\) is the vorticity, \(\gamma\) is the helicity, and \(\varphi=\arg(x+iy)\). The choices \(\gamma=0,\pi\) correspond to the two Néel helicities, whereas \(\gamma=\pm\pi/2\) correspond to the Bloch helicities. For this axisymmetric texture, \(N_{xy}(r)=(\varsigma/r)\sin\Theta(r)\,\partial_r\Theta(r)\), which is independent of \(\gamma\). With the skyrmion topological charge defined as \(Q=(4\pi)^{-1}\int d^2r\,N_{xy}\), the present boundary conditions give \(Q=-\varsigma\).

Since the magnetic texture varies slowly on the lattice scale, its effect on the itinerant electrons can be described by emergent gauge fields~\cite{PhysRevLett.121.097204,PhysRevB.110.054431,Zarezad_2024}. The opposite emergent fields on the two magnetic sublattices combine into effective spin-dependent, $\mathrm{s}=+ (\uparrow)/- (\downarrow)$, magnetic $\CMcal{B}_{\rm em}^{\mathrm{s}}
=\mathrm{s}\,\CMcal{B}_{\rm em}\hat{z}$, and electric $\CMcal{E}_{\rm em}^{\mathrm{s}}
=\mathrm{s}\,\CMcal{E}^{i}_{\rm em}\hat{e}_i$ gauge fields \cite{PhysRevLett.121.097204, SM},
where
$\CMcal{B}_{\rm em}(x,y)=-(\hbar/2e)\CMcal{N}_{xy}(x,y)$ and
$\CMcal{E}^{i}_{\rm em}(t;x,y)=-(\hbar/2e)\CMcal{N}_{ti}(t;x,y)$ are determined by the topological density
$\CMcal{N}_{\mu\nu}
=(\partial_\mu\n\times\partial_\nu\n)\!\cdot\!\n$ with $\mu,\nu \in \{t,x,y\}$ and $i \in \{x,y\}$.
Throughout this work, we consider a static skyrmion texture; hence, $\CMcal{E}^i_{\rm em}=0$, and focus on charge transport induced by the emergent magnetic field.

\textit{Boltzmann transport.}
To investigate the TOHE, we consider a strip geometry with periodic boundary conditions along the $x$ direction and a finite width $2w$ along the $y$ direction. The key quantity governing the topological contribution to the orbital Hall response is the nonequilibrium carrier distribution induced by the emergent magnetic field of the AFM skyrmion. We therefore determine the spin-dependent nonequilibrium distribution function $f_{\mathrm{s}}(\rv,\q)$ using the semiclassical Boltzmann equation~\cite{PhysRevB.82.184423,PhysRevB.97.134401},
\begin{equation}
\begin{aligned}
\resizebox{\linewidth}{!}{$\displaystyle
\v\cdot\frac{\partial f_{\mathrm{s}}}{\partial\rv}
-e\Big(\ev_{\rm ext}
+\v\times\CMcal{B}_{\rm em}^{\mathrm{s}}\Big)
\cdot
\frac{\partial f_{\mathrm{s}}}{\hbar\partial\q}
=
-\frac{f_{\mathrm{s}}-\langle f_{\mathrm{s}}\rangle}{\tau_{\mathrm{s}}}
-\frac{\langle f_{\mathrm{s}}\rangle-\langle f_{-\mathrm{s}}\rangle}{\tau_{\rm sf}},$}
\end{aligned}
\label{eq:Boltz}
\end{equation}
where $e > 0$ is the magnitude of the electron's charge, $\v$ is the electron velocity, $\langle ... \rangle$ denotes the angular average over momentum directions, and $\tau_{\mathrm{s}=\uparrow (\downarrow)}$ and $\tau_{\rm sf}$ denote the spin-dependent momentum and spin-flip relaxation times, respectively. Solving Eq.~(\ref{eq:Boltz}) to linear order in the emergent magnetic field for an external electric field $\ev_{\rm ext}=E_x\hat{x}$ yields the spin diffusion equation ~\cite{PhysRevB.110.054431}
\begin{equation}
\frac{d^2\overline{\delta\mu_z}(y)}{dy^2}
-
\frac{\overline{\delta\mu_z}(y)}{\lambda_{\rm sd}^2}
=
\left(\frac{e\tau E_x}{\widetilde{m}}\right)
\frac{d\overline{\CMcal{B}}_{\rm em}(y)}{dy}.
\label{eq:diff}
\end{equation}
Here, we define
$\delta\mu_z(x,y)=[\mu_\uparrow(x,y)-\mu_\downarrow(x,y)]/2$
as the local spin electrochemical-potential imbalance, with
$\mu_{\uparrow,\downarrow}(x,y)$ denoting the spin-resolved
electrochemical-potential shifts expressed in voltage units. For a strip of longitudinal period $L_x$, we define
$
\overline{F}(y)={L^{-1}_x}\int_{-L_x/2}^{L_x/2}F(x,y)\,dx$. For spin-dependent momentum-relaxation times $\tau_{s=\uparrow (\downarrow)}$, we define the spin-resolved diffusion length $\lambda_s^2=v_F^2\tau_s\tau_{\rm sf}/2$ and the spin diffusion length $1/\lambda_{\rm sd}^2=1/\lambda_\uparrow^2+1/\lambda_\downarrow^2$, where the Fermi velocity is $v_F=v_0\sqrt{1-P_{q_F}^{\,2}}$, $\mathrm{P}_{\q} = J/\abs{\varepsilon_\eta(\q)}$ parameterizes the overlap of electron wavefunctions of the two AFM sublattices, and $\tau=(\tau_\uparrow + \tau_\downarrow)/2$ is the spin-averaged momentum-relaxation time. We further define $\varepsilon_F$ and $q_F$ as the Fermi energy and the Fermi wave number, respectively, and  $\widetilde{m}=\varepsilon_F/v_0^2$ as the cyclotron effective mass.

Equation~(\ref{eq:diff}) shows that the gradient of the emergent magnetic field acts as the source of spin accumulation through the spin-dependent Lorentz force \cite{SM}. 
Equation~(\ref{eq:diff}) is solved subject to open boundary conditions at the sample edges $y=\pm w$, where the transverse spin-polarized current must vanish
$j^z_{y, \rm sp}(\pm w)=0$. The local spin imbalance is linearly proportional to the applied electric field $\overline{\delta\mu_z}(y) \propto E_x$.
The analytical solution satisfying open-boundary conditions is given in the SM~\cite{SM}.

Defining the spin-averaged electrochemical potential as
$\overline{\mu}(y)=[\overline{\mu}_{\uparrow}(y)
+\overline{\mu}_{\downarrow}(y)]/2$, we obtain its spatial profile from the open-circuit condition $j_{y,\mathrm{ch}}(y)=0$.
The integration constant is fixed by carrier-number conservation, $\int_{-w}^{w}\overline{\mu}(y)\,dy=0$,
which implies $\overline{\mu}(0)=0$ for the centered skyrmion. The resulting profile reads \cite{SM},
\begin{equation}
\overline{\mu}(y)
=
-p_\sigma\overline{\delta\mu_z}(y)
+
\frac{e\tau E_x}{\widetilde{m}}
(p_\sigma+p_\tau)
\int_{0}^{y}
\overline{\CMcal{B}}_{\rm em}(y')\,dy',
\label{eq:mu_profile}
\end{equation}
where $p_\tau=(\tau_\uparrow-\tau_\downarrow)/(\tau_\uparrow+\tau_\downarrow)$ and $p_\sigma=(\sigma_\uparrow-\sigma_\downarrow)/(\sigma_\uparrow+\sigma_\downarrow)$ parametrize the spin asymmetry of the momentum relaxation time and longitudinal charge conductivity, respectively; with $\sigma_{s=\uparrow(\downarrow)}$
denoting the spin-resolved longitudinal charge conductivity. 
For spin-degenerate carrier densities, the Drude relation gives $p_\sigma=p_\tau$. In an ideal $\mathcal{PT}$-symmetric antiferromagnet with symmetry-preserving disorder, $p_\tau=0$; a nonzero $p_\tau$ therefore represents a spin-selective scattering environment and is treated here phenomenologically. 
Both $\overline{\delta\mu}_z$ and $\overline{\mu}$ are
linear in $E_x$: the former controls the gradient-mediated
contribution to the local orbital accumulation, whereas the
latter generates the skyrmion-induced OHC correction below.

\textit{Total orbital Hall conductivity.}
The orbital Hall conductivity (OHC) associated with the intrinsic orbital Berry curvature is evaluated using the Kubo formalism with the Boltzmann nonequilibrium distribution~\cite{PhysRevB.106.104414},
\begin{equation}
\sigma^{z,\rm O}_{xy}
=
-g_{v}e\sum_{\eta,\mathrm{s}}
\int\frac{d^2\q}{(2\pi)^2}
f_{\eta\mathrm{s}}
\Omega^{z,\rm O}_{xy,\eta\mathrm{s}}(\q).
\label{ohc}
\end{equation}
Here, $g_v=2$ denotes the valley degeneracy, and the
momentum integral covers one valley. We consider
$\varepsilon_F>J$, with $f_{+,s}\equiv f_s$ obtained
from Eq.~(\ref{eq:Boltz}) and $f_{-,s}\simeq1$ describing the filled
valence band. The orbital Berry curvature is defined by
\begin{equation}
\Omega_{ij,\eta\mathrm{s}}^{z,\mathrm{O}}
=
2\hbar\,\mathrm{Im}
\sum_{\eta' \neq \eta}
\frac{
\bra{\Psi^{\mathrm{s}}_{\eta}}
\mathcal{J}^{z,\mathrm{O}}_{j}
\ket{\Psi^{\mathrm{s}}_{\eta'}}
\bra{\Psi^{\mathrm{s}}_{\eta'}}
v_{i}
\ket{\Psi^{\mathrm{s}}_{\eta}}
}{
(\varepsilon_{\eta'}-\varepsilon_{\eta})^2
},
\label{obc_cv}
\end{equation}
is the $z$ component of the orbital Berry curvature \cite{PhysRevLett.97.026603,PhysRevLett.99.197202,PhysRevB.105.195421}.
Because the velocity operator is diagonal in spin in the absence of spin-orbit coupling, only spin-conserving matrix elements with \(s'=s\) contribute to Eq. (\ref{obc_cv}).
Here, $\mathcal{J}^{z,\mathrm{O}}_{j}=(1/2)\{v_j,L^z\}$ is the orbital current operator, with the velocity operator $v_{i}=\hbar^{-1}\partial\mathcal{H}/\partial q_{i}$ and orbital angular momentum $L^z=-\hbar m^z/(g_L\mu_B)$, where $m^z$ is the $z$ component of the orbital magnetic moment, $g_L$ is the Land\'e $g$-factor for the orbital angular
momentum, and $\mu_B=e\hbar/(2m_e)$ is the Bohr magneton, with $m_e$ denoting the free-electron mass. 
The matrix elements of the orbital magnetic moment vector are evaluated using the modern theory of orbital magnetization~\cite{PhysRevLett.97.026603, PhysRevLett.99.197202, PhysRevB.105.195421},
\begin{equation}
\bm{m}_{\eta\mathrm{s},\eta'\mathrm{s}'}
=
-\frac{ie}{2\hbar}
\bra{\partial_{\bm{q}}\Psi^{\mathrm{s}}_{\eta}}
\times
\left[
{\mathcal{H}}
-
\frac{\varepsilon_{\eta}+\varepsilon_{\eta'}}{2}
\right]
\ket{\partial_{\bm{q}}\Psi^{\mathrm{s}'}_{\eta'}}.
\label{eq:omm}
\end{equation}
For the conduction band at low temperatures, we obtain
$
\Omega^{z,\rm O}_{xy,+\mathrm{s}}(\q)
=
-\Omega_0\,\mathrm{P}_{\q}^{5},
$
where $\Omega_0=e\hbar^3v_0^4/(4g_L\mu_{\rm B}J^3)$ \cite{SM}. 

Consequently, the total OHC can be decomposed into a constant linear-response OHC of the uniform AFM background, $\sigma^{z,\rm AFM}_{xy}$, and a position-dependent nonlinear topological contribution, $\Delta\sigma_{xy}^{z,\rm TOH}$, induced by the emergent gauge field of the AFM skyrmion,
\begin{subequations}\label{eq:ohc_mu}
\begin{align}
&{\sigma}^{z,\rm O}_{xy}(y)
=
\sigma^{z,\rm AFM}_{xy}
+ \Delta\sigma_{xy}^{z,\rm TOH}(y),
\label{eq:ohc_decomposition}
\\
& \sigma^{z,\rm AFM}_{xy}
=
\frac{-g_v e\Omega_0 J^2 \mathrm{P}_{\rm q_F}^{3}}
{3\pi\hbar^2v_0^2},
\label{eq:ohc_afm}
\\
&\Delta\sigma_{xy}^{z,\rm TOH}(y)=
\frac{-g_v e^2\Omega_0\varepsilon_{\rm F}\mathrm{P}_{\rm q_F}^{5}}
{\pi\hbar^2v_0^2}
\,\overline{\mu}(y).
\label{eq:ohc_toh}
\end{align}
\end{subequations}
The uniform linear-response OHC $\sigma_{xy}^{z,\mathrm{AFM}}$
includes contributions from the filled valence band and the occupied conduction-band states, whereas the skyrmion-induced contribution $\Delta\sigma_{xy}^{z,\mathrm{TOH}}$ arises from the nonequilibrium redistribution of carriers at the conduction-band Fermi surface.
A uniform linear OHC also occurs in nonmagnetic gapped
graphene~\cite{PhysRevB.103.195309}, but here the Dirac gap
originates from the $sd$ exchange coupling $J$. 
The total orbital Hall current \(j_y^{z,O}(y)=\sigma_{xy}^{z,O}(y)E_x\) contains the uniform linear-response background \(\sigma_{xy}^{z,\mathrm{AFM}}E_x\) and the skyrmion-induced topological contribution
\begin{equation}\label{jy}
\Delta j_{y}^{ z,\rm TOH}(y)=\Delta\sigma_{xy}^{z,\rm TOH}(y)E_x=\chi_{yxx}^{\mathrm{TOH}}(y) E_x^2,
\end{equation}
where we define the local second-order response \(\chi_{yxx}^{\mathrm{TOH}}(y)\equiv\Delta\sigma_{xy}^{z,\mathrm{TOH}}(y)/E_x\).  This establishes the nonlinear TOHE.

\begin{figure*}[t]
\centering
\includegraphics[width=0.33\textwidth]{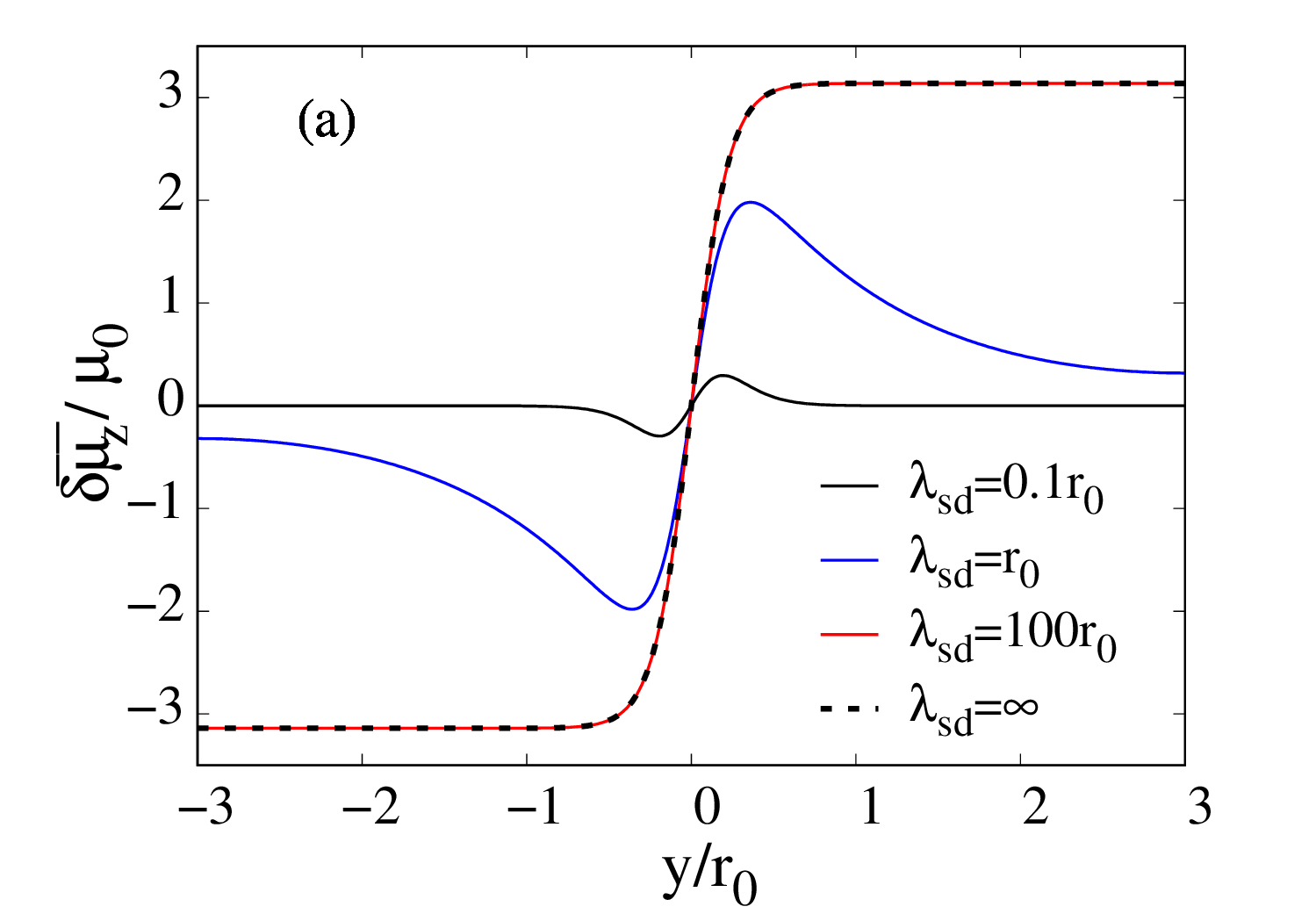}\hfill
\includegraphics[width=0.33\textwidth]{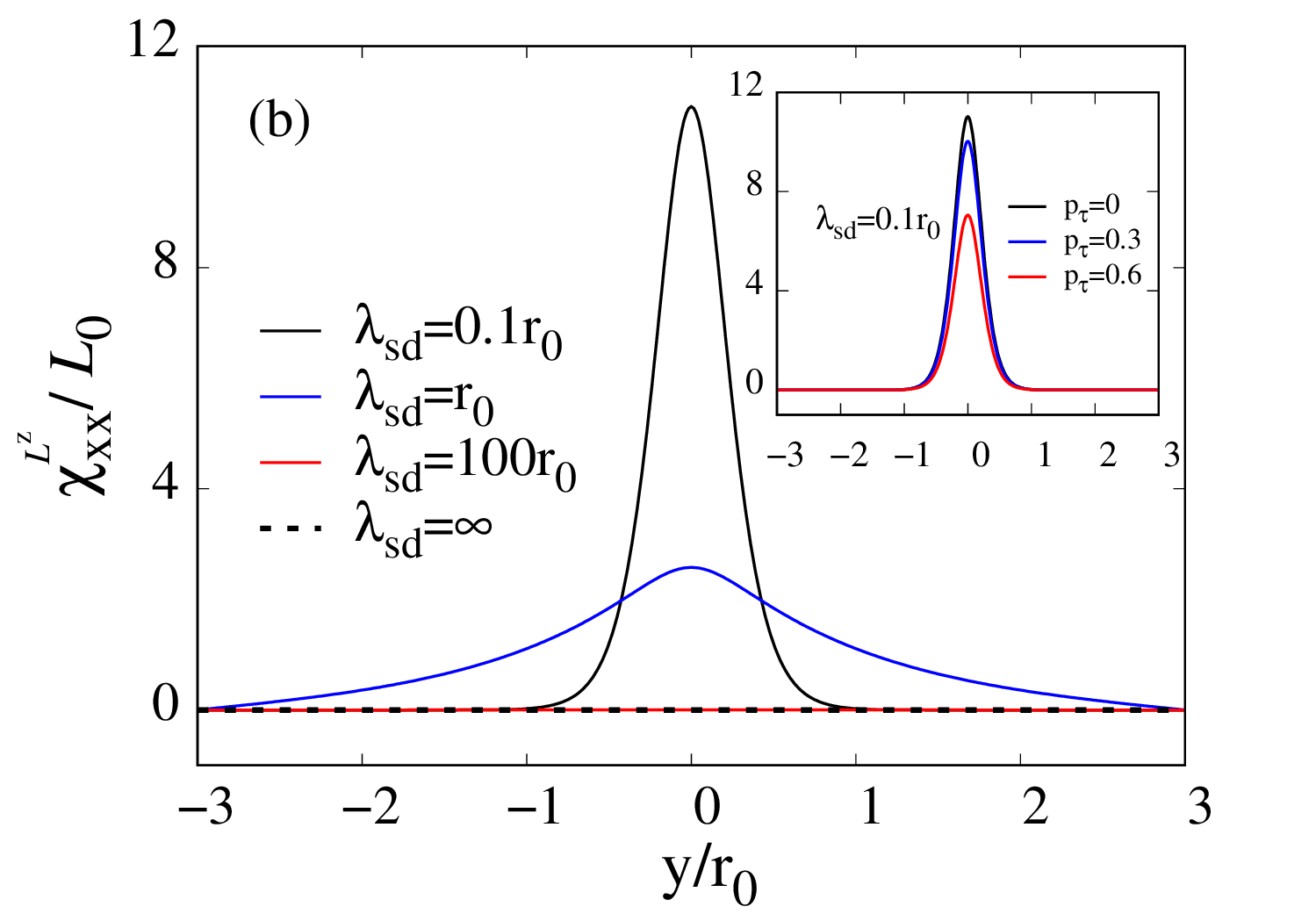}\hfill
\includegraphics[width=0.33\textwidth]{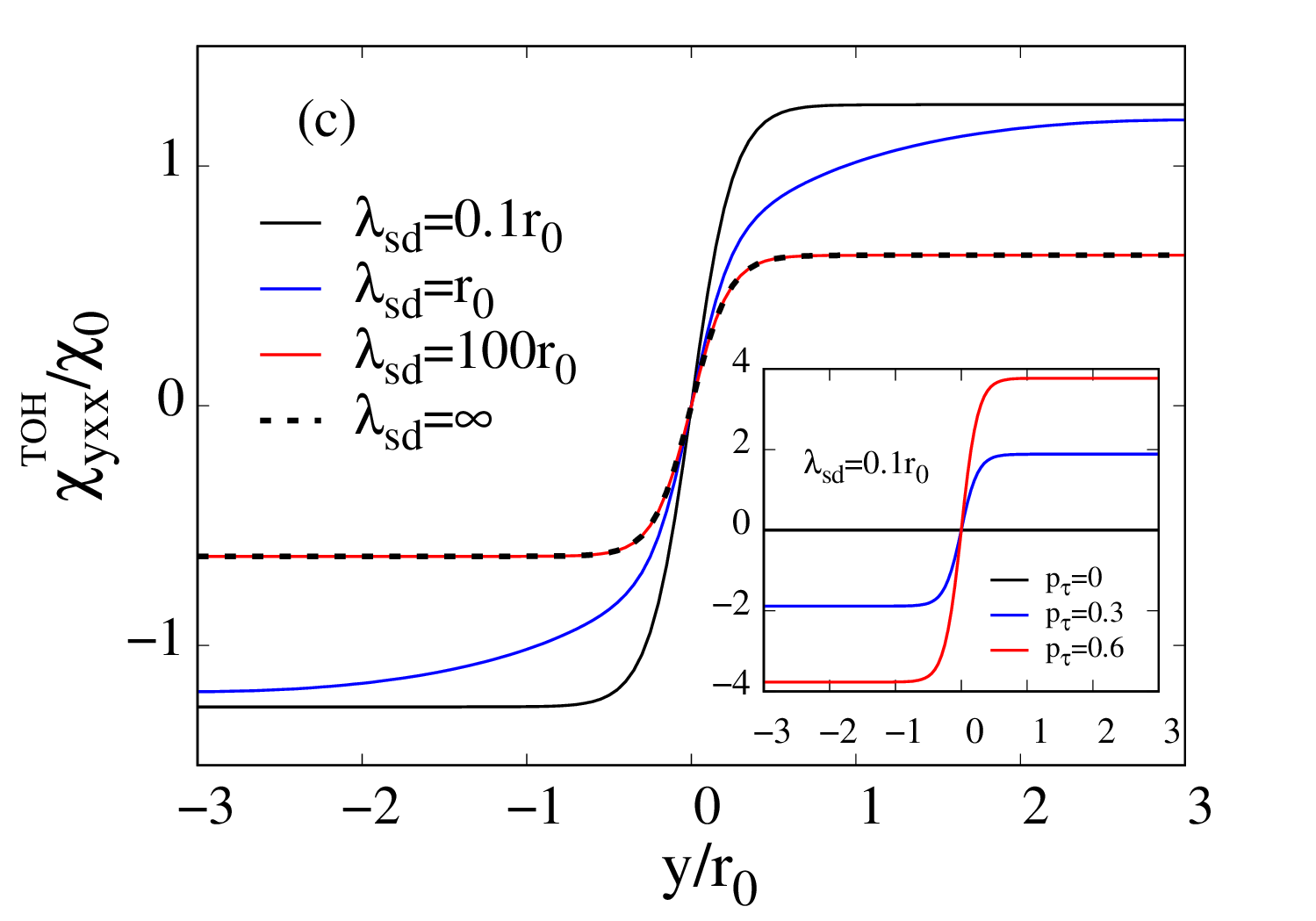}
\caption{Spatial profiles of (a) the spin electrochemical-potential imbalance, (b) the second-order orbital-accumulation coefficient and (c) the second-order topological orbital Hall coefficient, for different spin-diffusion lengths $\lambda_{\rm sd}$ and $p_\tau=0.2$. The insets in panels (b) and (c) show the corresponding quantities for different values of \(p_\tau\) at \(\lambda_{\rm sd}=0.1r_0\). The quantities are normalized by $\mu_0=e\tau v_0^2r_0B_0E_x/(16\varepsilon_{\rm F})$, $L_0=e^4v_0^4\hbar \tau^2 B_0\mathrm{P}_{\rm q_F}/(128\pi g_L\mu_{\rm B}\varepsilon_{\rm F}^3)$, and $\chi_0=-e^3\Omega_0\mathrm{P}_{\rm q_F}^5\tau  r_0B_0/(8\pi\hbar^2)$, where $B_0=(\pi r_0^2)^{-1}\int \CMcal{B}_{\mathrm{em}}dxdy$ is a characteristic emergent-field scale of the skyrmion. We set $L_x=8r_0$ and $w=3r_0$.}
\label{fig:results}
\end{figure*}

\textit{Orbital accumulation.}
Because orbital currents are often detected indirectly, we
also calculate the local nonequilibrium orbital accumulation
as a complementary observable. Using the nonequilibrium distribution obtained from the Boltzmann transport theory together with the Kubo formalism and the modern theory of orbital magnetization~\cite{PhysRevLett.97.026603, PhysRevLett.99.197202, PhysRevB.105.195421}, the induced orbital accumulation density can be decomposed into intraband and interband contributions,
$\delta\langle L^z\rangle
=
\delta\langle L^z\rangle^{\rm intra}
+
\delta\langle L^z\rangle^{\rm inter}$,
whose explicit expressions are provided in the SM \cite{SM}. For the present AFM model, the intraband contribution vanishes identically, leaving
\begin{equation}
\begin{aligned}
{\delta\langle L^z (y)\rangle}
&=\chi_{xx}^{L_z}E^2_x=
\frac{g_v e^2v_0^2\hbar}{g_L\mu_{\rm B}}
\frac{\mathrm{P}_{q_{\rm F}}}{16\pi\varepsilon_{\rm F}^2}
(1-p_\tau p_\sigma) \\
&\quad\times
\left[
-e\tau\frac{1}{E_x}\frac{d\overline{\delta\mu_z}}{dy}
+
\frac{(e\tau v_0)^2}{\varepsilon_{\rm F}}
\overline{\CMcal{B}}_{\rm em}
\right]E^2_x,
\end{aligned}
\label{eq:or_acc}
\end{equation}
where $\chi_{xx}^{L_z}$ is the local second-order orbital-accumulation coefficient.
Equation~(\ref{eq:or_acc}) contains a contribution mediated by the gradient of the spin imbalance and a direct emergent-field contribution. They partially cancel near the skyrmion center, while their relative signs depend on position. Their balance determines the local orbital accumulation.
To leading order in the emergent magnetic field, Eqs. ~\eqref{eq:diff},~\eqref{eq:mu_profile},~\eqref{eq:ohc_toh}, and~\eqref{eq:or_acc} imply \{\(\overline{\delta\mu_z},~\overline{}{\mu},~\Delta\sigma_{xy}^{z,\mathrm{TOH}}\} \propto \overline{\CMcal{B}}_{\rm em} E_x\), whereas \{\(\Delta j_y^{z,\mathrm{TOH}}\), \(\delta\langle L_z\rangle\} \propto \overline{\CMcal{B}}_{\rm em} E_x^2\). Consequently, the physical nonlinear current and orbital accumulation are even under \(E_x\rightarrow-E_x\) but odd under the reversal of the skyrmion topological charge, \(Q\rightarrow-Q\). In contrast, the uniform linear orbital Hall current is odd in \(E_x\) and independent of \(Q\).  
Since \(\mathcal{B}_{\mathrm{em}}\) is independent of the skyrmion helicity, $\gamma$, these nonlinear topological responses are also unchanged under helicity reversal within the present spin-orbit-free adiabatic model. 
This $Q$-odd parity contrasts with the quantum-regime TOHE of Ref.~\cite{Goebel2025}, for which the orbital Hall conductivity is unchanged under the reversal of the skyrmion topological charge.
An ac electric field, therefore, generates rectified dc and
second-harmonic orbital signals in this semiclassical regime.

\textit{Long spin-diffusion-length limit.}
In the limit $\lambda_{\rm sd}\gg w$, where $2w$ is the sample width, spin relaxation is negligible across the sample, and the spin electrochemical-potential imbalance satisfies \cite{SM};
\begin{equation}
	\frac{\overline{\delta \mu}_z(y)}{E_x}=\frac{e\tau v_0^2}{2\varepsilon_{F}}\int_{-w}^{w}d\tilde{y}\overline{\CMcal{B}}_{\mathrm{em}}(\tilde{y})\Big(\mathrm{H}(y-\tilde{y})-\mathrm{H}(\tilde{y}-y)\Big),
    \label{eq:delta_mu_nondiff}
	\end{equation}
where \(\mathrm{H}(y)\) denotes the Heaviside step function. The second-order TOH response becomes
\begin{align}\label{eq:ohc_nondiff}
\chi_{yxx}^{\mathrm{TOH}}(y)=
-\frac{g_v e^3\Omega_0\mathrm{P}_{\rm q_F}^5 \tau p_\tau }{\pi\hbar^2}
\int_{0}^{y}
\overline{\CMcal{B}}_{\rm em}(y')
\,dy'.
\end{align}
The orbital accumulation in this regime is obtained by substituting Eq.~\eqref{eq:delta_mu_nondiff} into Eq.~\eqref{eq:or_acc}. A careful evaluation of the derivative term shows that the two contributions in Eq.~\eqref{eq:or_acc} cancel exactly, yielding
\begin{equation}
\delta\langle L^z(y) \rangle = 0 \qquad (\lambda_{\rm sd} \to \infty).
\label{eq:or_acc_nondiff}
\end{equation}

Figure~\ref{fig:results} summarizes the central results of this work. 
Figure \ref{fig:results}(a) shows the transverse spin accumulation generated by the AFM skyrmion \cite{Zarezad_2024, PhysRevB.110.054431}, which provides the nonequilibrium input to the skyrmion-induced nonlinear orbital response. As the spin-diffusion length increases, the spin accumulation evolves toward a step-like cumulative profile that is constant outside the skyrmion core described by Eq.~(\ref{eq:delta_mu_nondiff}).

The second-order orbital-accumulation coefficient $\chi_{xx}^{L_z}(y)$ shown in Fig.~\ref{fig:results}(b) provides a potential spatially resolved signature of the nonlinear TOHE. In the diffusive regime, the orbital accumulation density exhibits a pronounced spatial profile that reflects the skyrmion texture. As \(\lambda_{\rm sd}\) increases, the accumulation density is progressively suppressed and vanishes when $\lambda_{\rm sd}\gg w$, as predicted by Eq.~\eqref{eq:or_acc_nondiff}.

Figure \ref{fig:results}(c) shows the local second-order topological orbital Hall coefficient \(\chi^{\mathrm{TOH}}_{yxx}(y)\). Its profile is odd in \(y\) and has zero width average in the symmetric strip. For \(p_\tau\neq0\), it approaches the finite profile in Eq. (\ref{eq:ohc_nondiff}) as \(\lambda_{\rm sd}/w\rightarrow\infty\).

The insets of Figs. \ref{fig:results}(b) and \ref{fig:results}(c) show complementary dependences on scattering asymmetry. At fixed \(\lambda_{\rm sd}\), with \(p_\sigma=p_\tau\), the Hall coefficient $\chi_{yxx}^{\rm TOH}$
is proportional to \(p_\tau\), whereas the local orbital accumulation is proportional to \(1-p_\tau^2\). Thus, the skyrmion-induced Hall-current correction vanishes for spin-symmetric scattering, while the local orbital accumulation can remain finite.

For an illustrative low-temperature parameter set with
$r_0=32~\mu\mathrm{m}$, $\lambda_{\rm sd}=4~\mu\mathrm{m}$,
and $p_\tau=p_\sigma=0$, a longitudinal field
$E_x=950~\mathrm{V/m}$ yields peak nonlinear orbital and linear
spin angular momentum densities of
$2.1\times10^{-4}\hbar\,\mu\mathrm{m}^{-2}$ and
$6.5\times10^{-4}\hbar\,\mu\mathrm{m}^{-2}$, respectively; see SM \cite{SM}.
At this field, the nonlinear orbital peak reaches approximately one-third of the linear spin peak at their respective spatial maxima, despite its quadratic field dependence. The localized orbital polarization motivates
spatially resolved measurements using orbital-sensitive
magneto-optical probes~\cite{PhysRevLett.131.156702}.
Quantitative detectability depends on the material-specific
magneto-optical coupling and orbital-relaxation length.

In summary, an isolated AFM skyrmion generates two
semiclassical second-order orbital observables distinct from
the uniform linear orbital Hall effect of the collinear
background: a skyrmion-induced orbital Hall-current
correction and a local orbital accumulation.
The local Hall-current correction is odd in \(y\) and has a zero width average in the symmetric strip, whereas the orbital accumulation is even.
Both arise
because the spin-dependent emergent Lorentz force reshapes
the nonequilibrium carrier distribution. The current
correction requires spin-asymmetric longitudinal scattering,
whereas the accumulation survives for
$p_\tau=p_\sigma=0$. As $\lambda_{\rm sd}/w$ increases,
the spin accumulation approaches a finite step-like profile
and the current correction remains finite for $p_\tau\neq0$,
while the local orbital accumulation vanishes because its
gradient-mediated and direct emergent-field contributions
cancel. Both nonlinear orbital signals are even under
$E_x\rightarrow-E_x$, odd under $Q\rightarrow-Q$, and
independent of the helicity $\gamma$ in the present
adiabatic, spin-orbit-free model. These signatures enable
rectified and spatially resolved detection of compensated
skyrmion topology and distinguish this mechanism from the
$Q$-even quantum-regime TOHE.

\emph{Acknowledgments---}
A.N.Z. thanks J. Abouie for his hospitality during a two-month visit to the Institute for Advanced Studies in Basic Sciences (IASBS), where this manuscript was prepared.
This work was supported by the Research Council of Norway through Grant Nos. 353919 and 361800 ``QTransMag'', and Grant No. 262633 ``QuSpin''.

\emph{Data Availability---}
Data supporting the findings of this study are contained within the article and Supplemental Material.

\bibliography{Report}

\end{document}